\documentclass[useAMS,usenatbib]{mn2e}
\usepackage{times,graphicx,epsfig}
\usepackage{float,lscape,color}
\usepackage{afterpage}
\usepackage{amssymb}
\usepackage{amsmath}
\def\PN{{\cal PN}}

\title[Dynamics of compact objects clusters: A post-Newtonian study] {Dynamics of
compact objects clusters: A post-Newtonian study} \author[G{\'a}bor Kupi, Pau
Amaro-Seoane and Rainer Spurzem]{G{\'a}bor Kupi$^{1}$, Pau Amaro-Seoane$^{2}$
and Rainer Spurzem$^{1}$\thanks{E-mail: kupiga@ari.uni-heidelberg.de (GK);
Pau.Amaro-Seoane@aei.mpg.de (PAS); spurzem@ari.uni-heidelberg.de (RS)}\\
$^{1}$Astronomisches Rechen-Institut, Zentr. Astron. Univ. Heidelberg (ZAH),
M\"onchhofstrasse 12-14, 69120 Heidelberg, Germany \\ $^{2}$Max-Planck Institut
f\"ur Gravitationsphysik (Albert-Einstein-Institut) Am M\"uhlenberg 1, D-14476
Potsdam, Germany} 

\date{Draft version of \today}

\pagerange{\pageref{firstpage}--\pageref{lastpage}} \pubyear{2006}

\begin{document}

\label{firstpage}

\maketitle

\begin{abstract}
Compact object clusters are likely to exist in the centre of some
galaxies because of mass segregation. The high densities and velocities reached
in them deserves a better understanding. 
The formation of binaries and their subsequent merging by
gravitational radiation emission is important to the evolution of such
clusters. We address the evolution of such a system in a relativistic regime. The
recurrent mergers at high velocities create an object with a mass much larger
than the average. 
For this aim we modified the direct {\sc Nbody6}++ code to include post-Newtonian
effects to the force during two-body encounters. We adjusted the equations of
motion to include for the first time the effects of both periastron shift and
energy loss by emission of gravitational waves and so to study the eventual
decay and merger of radiating binaries.
The method employed allows us to give here an
accurate post-Newtonian description of the formation of a run-away compact
object by successive mergers with surrounding particles, as well as the
distribution of characteristic eccentricities in the events. 
This study should be envisaged as a first step towards a detailed, accurate
study of possible gravitational waves sources thanks to the combination of the
direct {\sc Nbody} numerical tool with the implementation of post-Newtonian terms
on it.
\end{abstract}


\section{Introduction} 

It is nowadays well established that most, if not all, galaxies harbour a
supermassive black hole (SMBH) in their centre with a mass of some $10^{\,6-9}
M_{\odot}$ (see e.g. the recent review by \citealp{FerrareseFord2005,
Ferrarese2001,KormendyGebhardt2001}). There are also signs for masses of
$10^{\,6}M_{\odot}$ \citep{Greene2005}. In the case of our Galaxy this is even
imperative; an SMBH with a mass of about $\sim 3-4\times 10^6 M_{\odot}$
\citep{Eckart2002, Ghez2000, Ghez2003, Schoedel2002} must be ensconced in its
centre.  If one extends the correlation between the SMBH mass and the stellar
velocity dispersion of the bulge of the host galaxy (the ${\cal
M}_{\bullet}-\sigma$ correlation) observed for galactic nuclei
\citep{Gebhardt2000,FM00} to smaller systems, like globular clusters, one
should expect intermediate mass black holes (IMBH) with masses of between
$10^3-10^4 M_{\odot}$ to be lurking in the centres of such stellar clusters.
There are observations of M15 in the Milky Way or G1 in M31 \citep{Gerssen2002,
GebhardtRichHo2002,vanderMarel2003} which are compatible with this possibility,
but $N-$body models of these clusters have been made which do not require the
presence of an IMBH \citep{Baumgardt2003b}.

The densities observed in the central region of galaxies, where these very
massive objects are located, are very high and may even exceed the core density
of globular clusters by a factor hundred (about $10^7-10^8 M_{\odot}~{\rm
pc}^{-3}$ for the Galactic Centre, for instance) and thus make them very
special laboratories for stellar dynamics. 

On the other hand, it is not strictly excluded that the central mass
concentrations are not massive black holes (MBHs). Mass segregation creates a
flow of compact objects like neutron stars or stellar black holes to the
central parts of the cluster \citep{Lee1987,MEG00} and may constitute there a
cluster. This could mimic the effect of the MBH, and thus give an alternative
explanation of the properties of clusters that have gone core-collapse, like
M15 and G1 \citep{GebhardtRichHo2002, Baumgardt2003a,Baumgardt2003b,
vanderMarel2003}. On the other hand, MBHs are favoured in the case of
galaxies, in particular the Milky Way \citep{Maoz1998,Miller05}. 

For the case of a globular cluster it has been studied that stellar black holes
are probably ejected from the system. Stellar black holes should form three
body binaries and kick eachother out of the cluster \citep{Phinney1991,
Kulkarni1993, Sigurdsson1993, PortegiesZwart2000}. Nonetheless, if the velocity
dispersion is high enough, then binaries will not be created due to three body
encounters, as in the classical case considered before, but to gravitational
waves emission during two-body encounters. A simple way to understand this is
that the components of a binary merge before a third particle or a second
binary comes in sufficiently close to interact with them so as to eject the
binary or one of its. Thus, ejections cannot happen in such a scenario.  As a
matter of fact, for velocity dispersions of $\gtrsim 300$ km/s the merging time
in clusters with two mass components is already shorter than the required time
between interactions before a third particle or a second binary can bring about
an ejection \citep{HMLee1995}.

Relativistic stellar dynamics is of paramount importance for the study of a
number of subjects. For instance if we want to have a better understanding of
what the constraints on alternatives to supermassive black holes are; in order
to canvass the possibility of ruling out stellar clusters, one must do detailed
analysis of the dynamics of relativistic clusters and determine in particular
the core collapse time \citep{Miller05}. Also, when we want to more competently
dive into the formation of MBHs, learn how the dynamics around them is, for
instance to estimate captures of compact objects on a central SMBH via extreme
mass ratio inspiraling, or peruse a system of many supermassive black holes etc
the inclusion of relativistic effects is constitutive.

Our current work includes the study of stars on relativistic orbits around a
SMBH, so as to be able to estimate captures of compact objects on a central
SMBH via extreme mass ratio inspiraling and binary evolution of two SMBHs.

Efforts to understand the dynamical evolution of a stellar cluster in which
relativistic effects may be important have been already done by \cite{Lee1987},
\cite{QuinlanShapiro1989}, \cite{QuinlanShapiro1990} and \cite{MHL1993}. In his
work, \cite{MHL1993} (MHL93 from now onwards) addressed the problem of the
dynamical evolution of a cluster composed of compact objects by, with some
approximations, adding an estimate of the gravitational wave emission term
correction to {\sc Nbody5} (see section 3). Nevertheless, he neglected the
$1\PN~{\rm and}~2\PN$ terms and made use of the formalism introduced by
\citet{Peters1964}, possibly because of computational bourns. The computation
of the $\PN$ corrections is CPU-consuming, for we have to compute both, the
accelerations and their time-derivatives (see next section). Also, {\sc
Nbody5} is not suitable for supercomputers or special purpose GRAPE hardware;
here either {\sc Nbody6++} or {\sc Nbody4} have to be used \citep{Spurzem1999,Aarseth1999}.

In this work we describe a new tool that allows us to address this problem in a
much more rigurous way than done in the existing literature, including
deviations from the Newtonian formalism of the standard direct {\sc Nbody6++} code
\citep{Spurzem1999}, based on Aarseth's direct {\sc Nbody} codes
\citep{Aarseth1999}. We modified it in order to allow for post-Newtonian
($\PN$) effects, implementing in it the $1\PN,~2\PN~{\rm and}~2.5\PN$
corrections without any further approximation than those indwelling to the
calculation of the $\PN$ terms themselves \citep{Soffel1989}.  

In Section 2 we give a brief description of the method and of the
implementation of the $\PN$ terms into a standard {\sc Nbody} code. An
analysis of the formation and evolution of a particle that gains more and more
mass from successive mergers in the system (the ``runaway particle'') is made
in Section 3 and, to conclude, in Section 4 we make a summary and discussion of
the main results obtained.
\section{Method: Direct summation {\sc Nbody} with Post-Newtonian corrections}

The version of direct summation {\sc Nbody} method we employed for the calculations,
{\sc Nbody6}++, includes the {\em KS regularisation}. This means that when two
particles are tightly bound to each other or the separation among them becomes
smaller during a hyperbolic encounter, the couple becomes a candidate for a in
order to avoid problematical small individual time steps \citep{KS1965}. We
modified this scheme to allow for relativistic corrections to the Newtonian
forces by expanding the acceleration in a series of powers of $1/c$ in the
following way \citep{Damour1981,Soffel1989}:

\begin{eqnarray}
\underline{a}  = \underbrace{\underline{a}_0}_{\rm Newt.}
+\underbrace{\underbrace{c^{-2}\underline{a}_2}_{1\PN} +
\underbrace{c^{-4}\underline{a}_4}_{2\PN}}_{\rm periastron~shift} +
\underbrace{\underbrace{c^{-5}\underline{a}_5}_{2.5\PN}}_{\rm grav.~rad.} +
\mathcal{O}(c^{-6}),
\label{eq.a_expansionPN}
\end{eqnarray} 

\noindent where $\underline{a}$ is the acceleration of particle 1,
$\underline{a}_0= -{Gm_2\underline{n}}/{r^2}$ is the Newtonian acceleration,
$G$ is the gravitation constant, $m_1$ and $m_2$ are the masses of the two
particles, $r$ is the distance of the particles, $\underline{n}$ is the unit
vector pointing from particle 2 to particle 1, and the $1\PN,~2\PN~{\rm
and}~2.5\PN$ are post-Newtonian corrections to the Newtonian acceleration,
responsible for the pericenter shift ($1\PN,~2\PN$) and the quadrupole
gravitational radiation ($2.5\PN$), correspondingly, as shown in
Eq.\,(\ref{eq.a_expansionPN}). The expressions for the accelerations are:

\begin{eqnarray}
\begin{split}
&\underline{a}_2  =
\frac{Gm_2}{r^2}\mbox{\Huge\{}\underline{n}\Big[-v_1^2-2v_2^2+4v_1v_2+\frac{3}{2}(nv_2)^2+\\
&5\big(\frac{Gm_1}{r}\big)+4\big(\frac{Gm_2}{r}\big)\Big]+(\underline{v}_1-\underline{v}_2)
\big[4nv_1-3nv_2\big]\mbox{\Huge\}}
\label{eq.a2}
\end{split}
\end{eqnarray}

\begin{eqnarray}
 \begin{split}
&\underline{a}_4  = \frac{Gm_2}{r^2}\mbox{\Huge\{}\underline{n}
\mbox{\Huge[}-2v_2^4+4v_2^2(v_1v_2)-2(v_1v_2)^2\\
&+\frac{3}{2}v_1^2(nv_2)^2+\frac{9}{2}v_2^2(nv_2)^2-6(v_1v_2)(nv_2)^2\\
&-\frac{15}{8}(nv_2)^4+\big(\frac{Gm_1}{r}\big)\mbox{\Huge(}-\frac{15}{4}v_1^2+
\frac{5}{4}v_2^2-\frac{5}{2}v_1v_2\\
&+\frac{39}{2}(nv_1)^2-39(nv_1)(nv_2)+\frac{17}{2}(nv_2)^2\mbox{\Huge)}\\
&+\big(\frac{Gm_2}{r}\big)(4v_2^2-8v_1v_2+2(nv_1)^2\\
&-4(nv_1)(nv_2)-6(nv_2)^2)\mbox{\Huge]}\\
&+(\underline{v}_1-\underline{v}_2)\mbox{\Huge[}v^2_1(nv_2)+4v_2^2(nv_1)-5v_2^2(nv_2)\\
&-4(v_1v_2)(nv_1)+4(v_1v_2)(nv_2)-6(nv_1)(nv_2)^2\\
&+\frac{9}{2}(nv_2)^3+\big(\frac{Gm_1}{r}\big)\big(-\frac{63}{4}nv_1+\frac{55}{4}nv_2\big)\\
&+\big(\frac{Gm_2}{r}\big)\big(-2nv_1-2nv_2\big)\mbox{\Huge]}\mbox{\Huge\}}\\
&+\frac{G^3m_2}{r^4}\underline{n}\big[-\frac{57}{4}m_1^2-9m_2^2-\frac{69}{2}m_1m_2\big],
\end{split}
\label{eq.a4}
 \end{eqnarray}

\begin{eqnarray}
 \begin{split}
&\underline{a}_5   =
\frac{4}{5}\frac{G^2m_1m_2}{r^3}\mbox{\huge\{}(\underline{v}_1-
\underline{v}_2)\Big[-(\underline{v}_1-\underline{v}_2)^2+2\big(\frac{Gm_1}{r}\big)\\
&\quad\quad\quad\quad\quad\quad\quad\quad\quad\quad\quad\quad\quad\quad\quad\quad\quad
\quad\quad-8\big(\frac{Gm_2}{r}\big)\Big]\\
&+\underline{n}(nv_1-nv_2)\big[3(\underline{v}_1-\underline{v}_2)^2-6\big(\frac{Gm_1}{r}
\big)+\frac{52}{3}\big(\frac{Gm_2}{r}\big)\big]\mbox{\huge\}}.
\label{eq.a5}
\end{split}
\end{eqnarray}

\noindent 
In the last expressions $\underline{v}_1$ and $\underline{v}_2$ are the
velocities of the particles.  For simplification, we have denoted the vector
product of two vectors, $\underline{x}_1$ and $\underline{x}_2$, like
$x_{1}x_2$.  The basis of direct {\sc Nbody4} and {\sc Nbody6}++ codes relies
on an improved Hermit integrator scheme \citep{Makino1992, Aarseth1999} for
which we need not only the accelerations but also their time derivative. These
derivatives are not included in these pages for succinctness. We integrated our
correcting terms into the {\em KS regularisation} scheme as perturbations,
similarly to what is done to account for passing stars influencing a KS pair.
Note that formally the perturbation force in the KS formalism does not need to
be small compared to the two-body force \citep{Mikkola1997}. If the
internal KS time step is properly adjusted, the method will work even for
relativistic terms becoming comparable to the Newtonian force component.
\section{Dynamical evolution of a cluster of compact objects}

\subsection{The initial system and units}

The units used in our models correspond to the so-called $N$-body unit system,
in which $G=1$, the total initial mass of the stellar cluster is 1 and its
initial total energy is $-1/2$ \citep{Henon71a,HM86}.  The system was chosen to
be initially to be indentical to that employed by \cite{MHL1993}; i.e. a
spherical cluster with a number of compact stars ${\cal N}_{\star}=10^3$ of
identical mass $m$. These were distributed in an isotropic Plummer sphere,
which means that the phase-space distribution function is proportional to
$|E|^{7/2}$, where $E$ is the energy per unit mass of one star.  The density
profile is thus $\rho(r) = \rho_0{\left(1+\left({r}/{R_{\rm
Pl}}\right)^2\right)^{-5/2}}$, where $R_{\rm Pl}$ is the Plummer scaling
length.  For such a model the $N$-body length unit is $\mathcal{U}_\mathrm{l}
={16}/(3\pi)\,R_{\rm Pl}$.

In the situations considered here, the evolution of the cluster is driven by
2-body relaxation. A natural time scale is the (initial) {\em half-mass
relaxation time}. We use the definition of \citet{Spitzer87}, 

\begin{equation}
T_{\rm rh}(0) = \frac{0.138 N}{\ln \Lambda}
\left(\frac{R_{1/2}^3}{G{\cal M}_{\rm cl}}\right)^{1/2}.
\label{eq.trh}
\end{equation}

\noindent
For instance, for a Plummer model, the half-mass radius is $R_{1/2}
=0.769\,\mathcal{U}_\mathrm{l} = 1.305\, R_{\rm Pl}$.  ${\cal M}_{\rm cl}$ is
the total stellar mass and $\ln \Lambda = \ln\,(\gamma N)$ is the Coulomb
logarithm.

For the situation considered in this work, the square ratio of the central
velocity dispersion $\sigma_{\rm cen}$ to the speed of light $c$,

\begin{equation}
\Big(\frac{\sigma_{\rm cen}}{c}\Big)^2 \approx \frac{G {\cal M}_{\rm cl}}{{\cal
R}_{\rm cl} c^2} \approx \frac{{\cal R}_{\rm Schw}^{\rm cl}}{{\cal R}_{\rm cl}}
\label{eq.relativ_cond}
\end{equation}
 
\noindent
is big enough, so that we can expect that relativistic effects play a
noticeable role in the evolution of the system. For this aim, we chose
$\sigma_{\rm cen}$ to be $\sim 4300$ km/s. $G$ is the gravitational constant,
${\cal R}_{\rm cl}$ is the radius of the cluster and ${\cal R}_{\rm Schw}^{\rm
cl}= 2 G {\cal M}^{\rm cl}/c^2$ is the Schwarzschild radius of the cluster.

In our calculations the $\PN$ terms are acting all the time during the
calculations but obviously become important only when velocities are high.  Our
criterion for particles to merge is that they reach their common
Schwarzschild radii ${\cal R}_{\rm Schw}$; i.e. the sum of their Schwarzschild
radii. This is of course approximative because the $\PN$ treatment breaks down
when particles are that close (and $v \sim c$), but this should not matter, for
the merging phase is {\em much} faster than any stellar dynamical time.  The
gravitational recoil, the expected lose linear momentum in asymmetric systems
in which the merger remnant receives a kick from the gravitational waves
emission obviously does not show up in our models, because we truncate the
series at $\mathcal{O}(c^{-5})$ and it is only to be treated as an effect of
higher-order terms.

\subsection{Formation of a run-away body}

\begin{figure}
\resizebox{\hsize}{!}{\includegraphics[clip]
{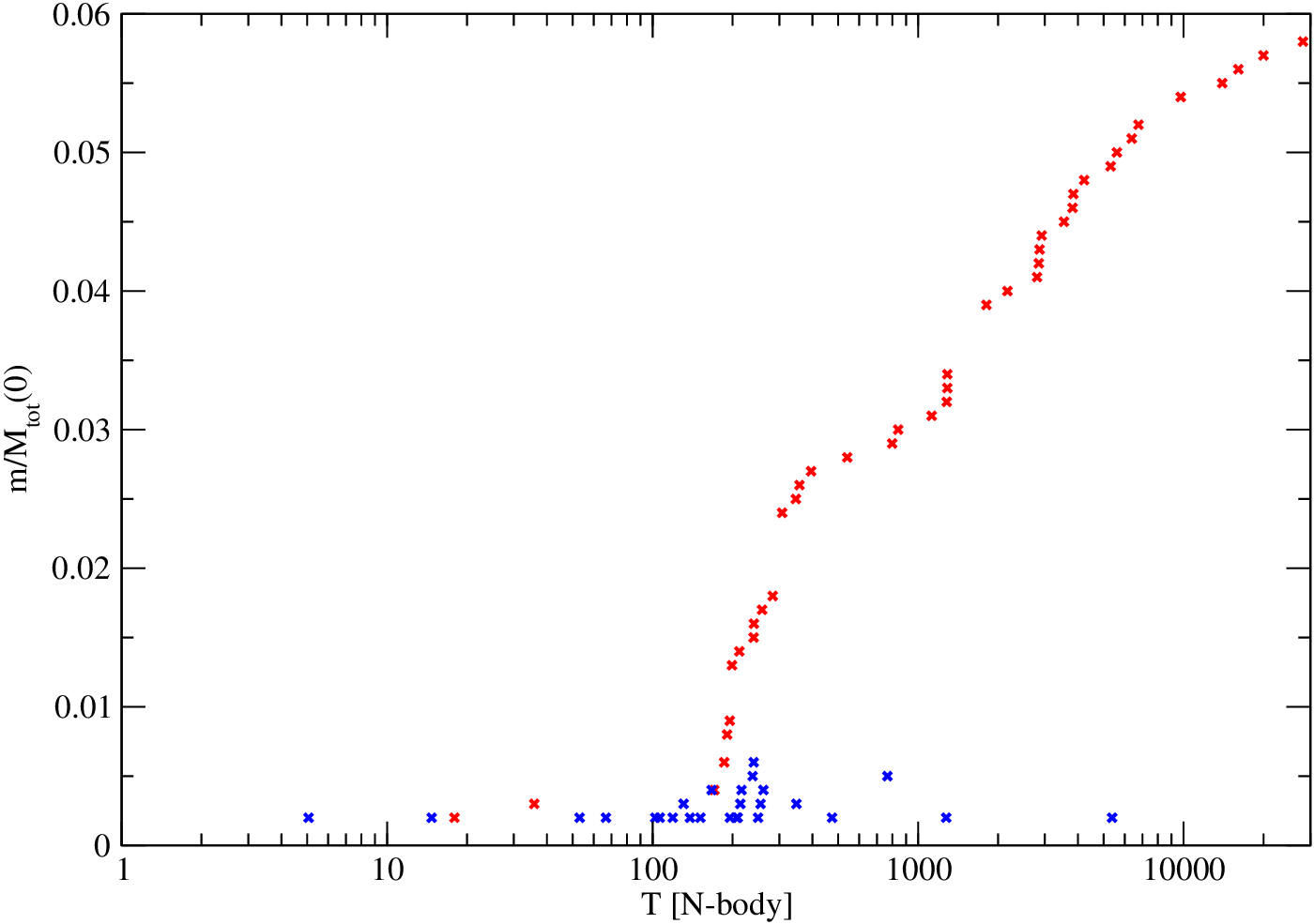}}
\caption{Time evolution of merging masses. The formation of the runaway
particle is about the time of the cluster core collapse. Fore more details
see text
\label{fig.m}
}
\end{figure}

Even though we started with a single-mass stellar system, the masses of some
objects in the cluster increased by relativistic mergers. In
Fig.\,(\ref{fig.m}) we survey the time evolution of the mass increase. We find
a number of mergers that lead to the variation of the initial single mass
situation. The particle masses increase after the relativistic merging events,
since we are assuming that the particles merge perfectly when they reach the
distance of their ${\cal R}_{\rm Schw}$ (see above). We find the formation of a
runaway particle that reaches almost six percent of the initial total mass by
the end of the simulation, see Fig.\,(\ref{fig.m}).  We denoted the mass of
runaway body by red crosses and the mass of other mergers by blue crosses. 

\begin{figure}
\resizebox{\hsize}{!}{\includegraphics[clip]
{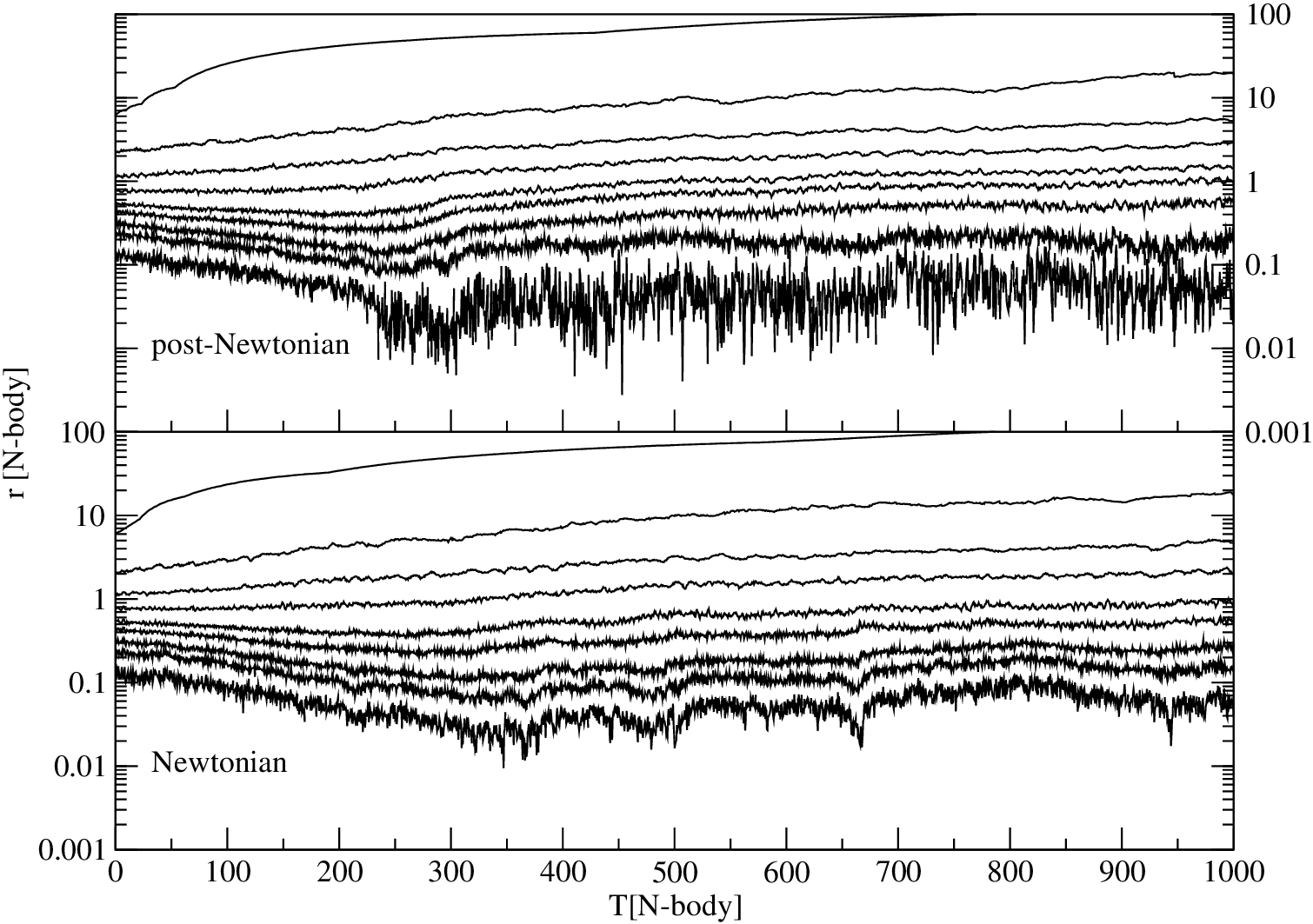}}
\caption{Evolution of the Lagrangian radii corresponding from the bottom to the top 
to 1\%, 5\%, 10\%, 20\%, 30\%, 50\%, 70\%, 90\% and
100\% of the total mass
\label{fig.Lagr}
}
\end{figure}

One can observe that the runaway body dominates the system after its fast
growing phase around 300 time units, which is approximately the moment at which
the core collapse of the system happens, as we can see in
Fig.\,(\ref{fig.Lagr}).  Only some merger events which are independent from the
runaway body can occur after this phase.  This fast growing phase occurs at the
core collapse of the system \citep{MH97}. In Fig.\,(\ref{fig.Lagr}) we follow
the evolution of the so-called Lagrangian radii of the system, spheres
containing 1\%, 5\%, 10\%, 20\%, 30\%, 50\%, 70\%, 90\% and 100\% of the total
mass of the cluster; the centre of the cluster is defined to be the centre of
the mass density. Since the runaway particle is included, and in the end its
mass reaches 5\% of the total initial mass of the cluster, the curves
corresponding to 1\% and 5\% roughly correspond to its evolution. We observe
that the runaway stops the core collapse and allows for a expansion.  

The process of mergers translates directely into a production of energy in the
central regions of the cluster. The centre adapts to supply the cluster with
the same amount of energy that it can obtain via relaxation, and this amount is
determined by the large-scale structure.


According to, for instance, the table given in \cite{FB01}, the standard value
for the core collapse time is of roughly $\sim 15-20$ times the half-mass
relaxation time $T_{\rm rh}$.  We find nevertheless that the core collapse time
is $t_{\rm cc} \sim 11 T_{\rm rh}$, with a value of $\gamma=0.11$ in the
Coulomb logarithm \citep{GH94}, which clearly suggests that the ${\PN}$ terms
accelerate the collapse.  This can be seen more clearly in
Fig.\,(\ref{fig.Lagr}), which corresponds to the same simulation but without
making use of relativistic corrections. There we can see that $t_{\rm cc} \sim
380 \sim 14 T_{\rm rh}$.

\begin{figure}
\resizebox{\hsize}{!}{\includegraphics[clip]
{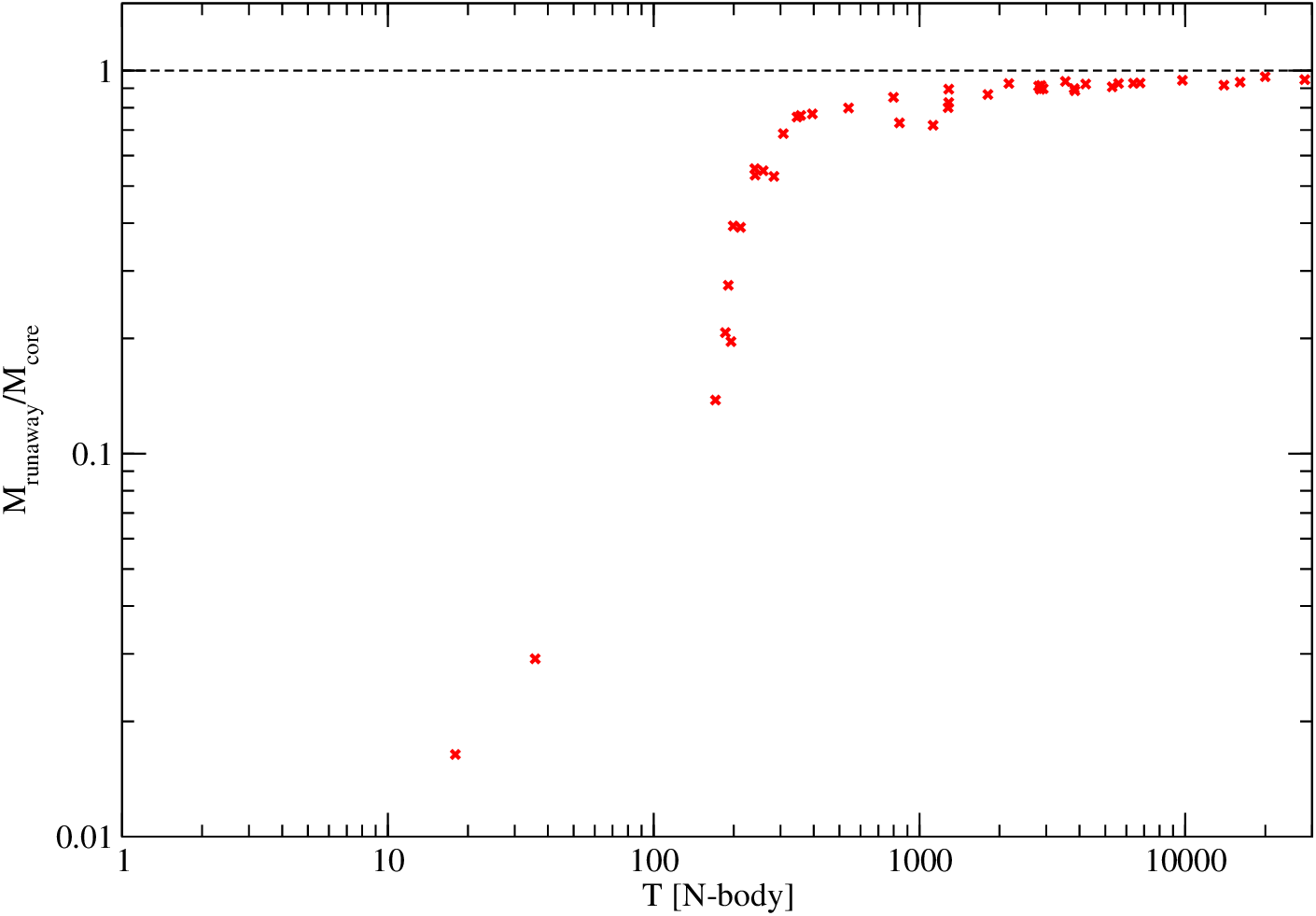}}
\caption{Evolution of the runaway particle mass in units of the core mass (at the same time)
\label{fig.Mrunaway}
}
\end{figure}

In Fig.\,(\ref{fig.Mrunaway}) we show the evolution of the the runway particle
mass normalised to the mass contained in the core of the cluster, defined as in
\cite{CH85}.  The mass of the runway particle can grow only up to the core
mass. The core mass continuously decreases as the core collapse proceeds. We
see this in the figure, where the runaway particle grows and saturates to the
core mass after $\sim 1200$ time units.

The evolution and formation of the runaway particle mass is not as fast as it
was in MHL93, as we can see in his Figure 5. For our simulation the sudden
jump in the growth of the mass comes in slightly later and is smoother,
reaching final values for the runaway particle mass of about three times
smaller than in MHL93. The reasons for the differences are to be attributed to
the following: MHL93 calculated the influence of the $2.5\PN$ term on the
orbits in an unperturbed pair and made them merge after a {\em decay
timescale}, following the \cite{Peters1964} formalism. This requires the
assumption that particles move along their orbits on an ellipsis, only valid
when they are very far from the relativistic regime. On the other hand, we
implemented the $2.5\PN$ term in the code itself, so that the relativistic
corrections are a natural feature whose influence on the evolution of the
system comes in when the velocities of the stars become high enough.  The
influence of the $1\PN$ and $2\PN$ terms, corresponds to the conservative phase
evolution of the orbit and cannot be relevant because they do not change its
energy and angular momentum.

\section{Conclusions}

In this work we have presented a study of the formation and evolution of a
runaway particle in a dense cluster of compact objects -which initially had the
same mass- as a result of relativistic mergers. We employed a modified version
of the direct summation {\sc Nbody6} code in which we have implemented the
$1\,\PN$, $2\,\PN$, and $2.5\,\PN$ terms to take into account post-Newtonian
corrections to the standard {\sc Nbody} Newtonian acceleration. 

The runaway particle reaches in the end of our simulations $\sim\,6\%$ of the
initial total stellar mass of the cluster. We have also compared our work to
a previous result based on a more approximative scheme, the approach
described in \cite{Peters1964} and we have found out that the net result is
that the growth of the runaway particle in the study of MHL93 is $\sim$ 3 times larger.
Since the $1\,\PN$, $2\,\PN$ terms modify the extrinsic features of the orbits
(e.g. the orientation) but do not affect their intrinsic parameters (like
frequency), we therefore can expect their effect to be averaged out during the
evolution of the system and not influence the mergers rates. One should thus
attribute the differences to the approach he made, somehow inadequate for the
velocity regime considered.

This study should be envisaged as successful test test of the code, which
showed to be robust. This tool can be applied to other astrophysical scenarios
which require a post-Newtonian treatment. This includes on-going work, as e.g.
captures of compact objects by a supermassive black hole in a galactic centre,
also known as extreme mass ratio inspirals (EMRIs).  One of the fundamental
aims is to rigurously explore the parameter space, so that we can provide the
LISA data analysis community with realistic estimates of, for instance, the
eccentricity, mass ratio etc at the beginning of the final merger, when the
smaller compact object enters the LISA band. An assumption for the initial
parameter space is necessary in order to develop waveform "banks" for this kind
of events.  One must note here that the inclusion of the $1\,\PN$ and $2\,\PN$
terms is very relevant, for ressonant relaxation (or Kozai) effects, which may
increase the rate of inspiral significantly, may be strongly affected by by
relativistic precession and thus have an impact on the number of captures
\citep{HopmanAlexander06,Kozai62}.
The inclusion of higher-order $\PN$ terms is also part of current study and
will also shed light on other aspects of this subject (spin-spin coupling,
spin-orbit interaction and radiation recoil).
\section*{Acknowledgments}

We acknowledge financial support, also for computing resources (GRACE cluster)
of the Volkswagen Foundation, SFB439 at the Univ. of Heidelberg, and the State
of Baden-W{\"u}rttemberg.  

\noindent
GK is thankful to the Max-Planck Institut f{\"u}r Gravitationsphysik for his
visit during the month of March.  

\noindent
RS thanks A.  Gopakumar and G. Sch{\"a}fer for many enlightening discussions.  

\noindent
The work of PAS has been supported partially by the SFB439 project at the ARI
in Heidelberg and in the framework of the Third Level Agreement between the DFG
(Deutsche Forschungsgemeinschaft) and the IAC (Instituto de Astrof\'\i sica de
Canarias) at AEI in Potsdam. PAS would like to thank Curt Cutler for discussing
on the difficult implementation of the $\PN$ terms into the analysis and Marc
Freitag for general comments on the main goals of the study and useful remarks
during his redaction of the paper.  PAS would also thank Leor Barack for his
impressions on a first draft. 

\noindent
We are indebted with the referee, H.M. Lee, for comments that led to an
improvement of the work.
%

    %

\label{lastpage}

\end{document}